\documentclass[fleqn,usenatbib,useAMS]{mnras}

\usepackage{graphicx}	% Including figure files
\usepackage{amsmath}	% Advanced maths commands
\usepackage{amssymb}	% Extra maths symbols
\usepackage{multicol}        % Multi-column entries in tables
\usepackage{bm}		% Bold maths symbols, including upright Greek
\usepackage{pdflscape}	% Landscape pages

\usepackage[T1]{fontenc}
\usepackage{ae,aecompl}

\usepackage{newtxtext,newtxmath}
\title[Collisionless shocks with anisotropic upstream]{Density jump as a function of magnetic field strength for perpendicular collisionless shocks with anisotropic upstream pressure}

\author[A. Bret et al.]{
Antoine Bret$^{1,2}$\thanks{E-mail: antoineclaude.bret@uclm.es}
\\
$^{1}$ETSI Industriales, Universidad de Castilla-La Mancha, 13071 Ciudad Real, Spain\\
$^{2}$Instituto de Investigaciones Energ\'{e}ticas y Aplicaciones Industriales, Campus Universitario de Ciudad Real, 13071 Ciudad Real, Spain
}

\date{Last updated -; in original form -}

\pubyear{2023}

\begin{document}
\label{firstpage}
\pagerange{\pageref{firstpage}--\pageref{lastpage}}
\maketitle

\begin{abstract}
Shock waves are common in astrophysical environments. On many occasions, they are collisionless, which means they occur in settings where the mean free path is much larger than the dimensions of the system. For this very reason, magnetohydrodynamic (MHD) is not equipped to deal with such shocks, be it because it assumes binary collisions, hence temperature isotropy, when such isotropy is not guaranteed in the absence of collisions. Here we solve a model capable of dealing with perpendicular shocks with anisotropic upstream pressure. The system of MHD conservation equations is closed assuming the temperature normal to the flow is conserved at the crossing of the shock front. In the strong shock sonic limit, the behavior of a perpendicular shock with isotropic upstream is retrieved, regardless of the upstream anisotropy. Generally speaking, a rich variety of behaviors is found, inaccessible to MHD, depending on the upstream parameters. The present work can be viewed as the companion paper of MNRAS 520, 6083–6090 (2023), where the case of a parallel shock was treated. Differences and similarities with the present case are discussed.
\end{abstract}


%%%%%%%%%%%%%%%%%%%%%%%%%%%%%%%%%%%%%%%%%%%%%%%%%%

%%%%%%%%%%%%%%%%% BODY OF PAPER %%%%%%%%%%%%%%%%%%

% The MNRAS class isn't designed to include a table of contents, but for this document one is useful.
% I therefore have to do some kludging to make it work without masses of blank space.
%\begingroup
%\let\clearpage\relax
%\tableofcontents
%\endgroup
%\newpage

\section{Introduction}
Shock waves are extremely frequent in astrophysical environments. They are present in the solar system \citep{burgess2015}, in supernova remnants \citep{McKee1980} and a key ingredient of some scenarios for cosmic rays generation \citep{Blandford78,Bell1978a,Bell1978b}, gamma-ray bursts \citep{Piran2004} or fast radio bursts \citep{Zhang2020}.

It turns out that many of these shockwaves are collisionless, which means that they occur in settings where the mean free path is much larger than the dimensions of the system. Consequently, instead of being mediated by binary collisions, they are mediated by collective plasma effects \citep{Sagdeev66,sagdeev1991}.

Formally speaking, analyzing such shocks would require working at the kinetic level using the Vlasov equation. However, this partial derivative equation has proved extremely difficult to deal with. As an illustration of this point, the Fields Medal was awarded in 2010 to Cedric Villani for solving the problem of nonlinear Landau damping \citep{Mouhot2011,VillaniPoP}. In view of such a mathematical challenge, magnetohydrodynamic (MHD) is frequently used to analyze collisionless shocks, even though MHD eventually relies on fluid dynamics which assumes a small Knudsen number, that is, a small mean free path ensuring pressure isotropy (\cite{Goedbloed2010}, chaps. 2 and 3 or \cite{TB2017}, section 13.2).

An important feature of shockwaves, whether they are collisional or not, is their density jump. In the case of a collisionless plasma, the absence of binary collisions may result in long-lasting pressure anisotropies. These can in turn trigger important departures from the MHD behavior which, by default, assumes isotropic pressures.

A notable change in density jump can have important physical consequences. For example, it is well known that the index of shock accelerated particles scales like $(r-1)^{-1}$, where $r$ is the shock density ratio \citep{Blandford78}. Hence, a lower $r$ results in a steeper cosmic ray spectrum.

In the absence of an ambient magnetic field, an anisotropic collisionless plasma is Weibel unstable \citep{Weibel,SilvaPRE2021}, so that using the MHD jump conditions to analyze unmagnetized collisionless shocks can be justified. Indeed, in this case, collisionless instabilities eventually play the role of binary collisions, as was also found for the  mechanism of collisionless shock formation \citep{BretPoP2013,Bret2015ApJL}.

Yet, it is well known that an ambient magnetic field can stabilize an anisotropy in a collisionless plasma \citep{Hasegawa1975,Gary1993}. This has been beautifully demonstrated by \emph{in situ} measurement in the solar wind \citep{BalePRL2009,MarucaPRL2011,SchlickeiserPRL2011}. In such cases, the density jump of a collisionless shock may depart strongly from its MHD predicted behavior.

A stable anisotropy may develop in various ways in a collisionless shock. Even when starting from an isotropic upstream, the crossing of the front can trigger an anisotropy stable in the downstream. In addition, the upstream itself can be isotropic, as has been measured on occasions in the solar system \citep{Fraschetti2020}. In such a case, the isotropy of the upstream may very well makes its way to the downstream.

Whether a pressure isotropy arises from a disturbance when crossing the shock front, or from an anisotropic upstream, or from both, MHD is not equipped to fully deal with such situations. Various authors derived MHD jump conditions accounting for anisotropic pressures \citep{Karimabadi95,Erkaev2000,Vogl2001,Gerbig2011}. But even though in these equations the upstream anisotropy can be considered an input, the downstream isotropy is left as a free parameter. As will be checked later in this article, there are simply not enough MHD equations to constrain it.

In a series of recent works \citep{BretJPP2018,BretPoP2019,BretPoP2021,BretJPP2022,BretJPP2022a,Haggerty2022,BretMNRAS2023}, a method was developed capable of accounting for anisotropies when computing the density jump of a magnetized collisionless shock. It was found, for example, that for a strong sonic parallel shock, whereas MHD predicts a density jump of 4, a strong enough magnetic field can bring it down to 2 \citep{BretJPP2018}. This theoretical prediction was successfully checked by particle-in-cell (PIC) simulations \citep{Haggerty2022}.

So far, our model is limited to pair plasmas. While pair plasmas are definitely present in some astrophysical environments \citep{Siegert2016}, the main motivation to choose pair plasma is simplicity: Dealing with anisotropies demands considering by default different temperatures in the directions parallel and perpendicular to the field. In a pair plasma, these temperatures will be the same for electrons and positrons. In an electron/ion plasma, they won't \citep{Guo2017,Guo2018}, resulting in a model with four different temperatures instead of two. Note however that some preliminary results show that the conclusion derived for parallel shocks in pair plasmas are equally valid for parallel shock in electron/ion plasma \citep{Shalaby2023}.

Shocks waves, colisionless or not, can propagate in the direction parallel or perpendicular to the ambient magnetic field. The former are called ``parallel shocks'', the latter, ``perpendicular shocks''. In \cite{BretMNRAS2023}, the case of a parallel collisionless shock with an anisotropic upstream has been treated. Here we treat the case of a perpendicular shock, still with an anisotropic upstream.

The paper is structured as follows: the method used to close the system of MHD conservation equations in the presence of a pressure anisotropy is explained in Section \ref{sec:method}. Then the results of the model are directly presented in Section \ref{sec:results}, with the details of the algebra reported in various appendices.

\begin{figure}
\begin{center}
 \includegraphics[width=\columnwidth]{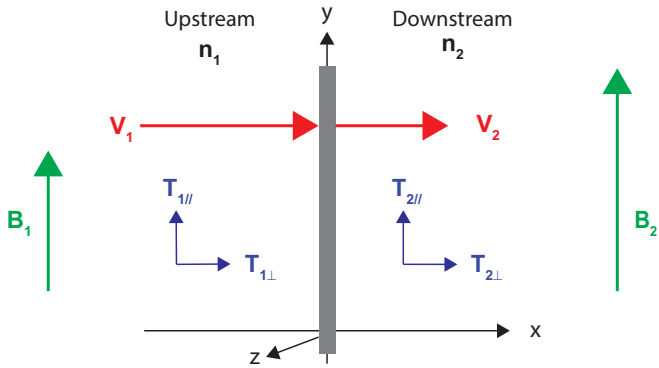}
\end{center}
\caption{System considered. The field is perpendicular to the flow in the upstream and the downstream. The directions ``parallel'' and ``perpendicular'' are defined with respect to the field.}\label{fig:system}
\end{figure}

\section{Conservation equations and method}\label{sec:method}
The system considered is pictured on figure \ref{fig:system}. The plasma goes from left to right. Upstream quantities have subindexes ``1’’ and downstream quantities subindexes ``2’’. The field is normal to the flow in both the upstream and the downstream \citep{Kulsrud2005}. For the temperatures, the directions ``parallel’’ and ``perpendicular’’ are in relation to the local magnetic field.

The MHD jump conditions for anisotropic temperatures have been derived by several authors \citep{Hudson1970,BretJPP2022}. For  an adiabatic index of $\gamma = 5/3$ they read,
  \begin{eqnarray}
 \left[n_i v_i\right]_1^2&=& 0, \label{eq:conser1} \\
 \left[n_i k_B T_{\perp i} +\frac{B_i^2}{8 \pi }+m n_i v_i^2\right]_1^2 &=& 0, \label{eq:conser2} \\
 \left[B_i v_i\right]_1^2&=& 0,  \label{eq:conser3} \\
\left[\frac{n_i v_i}{2} (k_B T_{\parallel i}+4k_B T_{\perp i}) + n_i v_i\frac{1}{2} m v_i^2 +   v_i\frac{ B_i^2}{4 \pi }\right]_1^2 &=& 0, \label{eq:conser4}
 \end{eqnarray}
where $[Q_i]_1^2$ denotes the difference $Q_2-Q_1$ and $k_B$ the Boltzmann constant. These equations will be analyzed and solved in terms of the dimensionless variables,
\begin{eqnarray}\label{eqs:dimless}
 r     &=& \frac{n_2}{n_1},   \\
 A_i   &=& \frac{T_{\perp i}}{T_{\parallel i}},   \nonumber\\
\sigma &=& \frac{\frac{1}{2} m n_1  v_1^2}{B_1^2/8 \pi } , \nonumber\\
\chi_1^2 &=& \frac{mv_1^2}{k_BT_{\perp 1}}.   \nonumber
\end{eqnarray}
The parameter $\chi_1$ is very close to the upstream Mach number. Yet, the present model constraints the degrees of freedom of the system, resulting in a varying effective adiabatic index \citep{BretPoP2021}. It is therefore preferable to leave it out of this parameter.

The $\sigma$ parameter is closely related to the Alfv\'{e}n Mach number $M_A$ through,
\begin{equation}\label{eq:MA}
M_A = \frac{1}{\sqrt{\sigma}}.
\end{equation}
Yet, conducting the analysis in terms of $\sigma$ allows for a direct comparison with the results of \cite{BretMNRAS2023}, and with PIC simulations of collisions shocks which usually parameterize the field in terms of $\sigma$ (see for example \cite{Sironi2011ApJ}).

Considering upstream quantities as inputs, these 4 equations contain 5 unknowns: $n_2$, $v_2$, $B_2$, $T_{\parallel 2}$ and $T_{\perp 2}$. One more equation is missing that MHD cannot provide, which is why previous analysis of Eqs. (\ref{eq:conser1}-\ref{eq:conser4}) considered the downstream anisotropy $T_{\perp 2}/T_{\parallel 2}$ a free parameter \citep{Karimabadi95,Erkaev2000,Vogl2001,Gerbig2011}.

In \cite{BretJPP2018} we proposed an \emph{ansatz} for a fifth equation : as the plasma crosses the  front, its temperature perpendicular to the motion is conserved. This intuitively stems from the fact that the plasma is compressed in the direction parallel to the motion. Not in the perpendicular one. This hypothesis has been successfully tested by PIC simulations \citep{Haggerty2022}.

In the present case, this prescription bears with the temperatures parallel to the field, since the field is perpendicular to the motion. Therefore, the fifth equation allowing to close the system (\ref{eq:conser1}-\ref{eq:conser4}) is simply,
\begin{equation}\label{eq:fifth}
 T_{\parallel 2}=T_{\parallel 1}.
\end{equation}
Now, the resulting upstream can be stable or unstable. More precisely, it can be firehose or mirror unstable (see Section \ref{sec:insta}). If the field $\bmath{B}_2$ is strong enough to stabilise it, then this is the end state of the downstream. If not, the downstream migrates to its marginal stability threefold, reaching the end state.

\begin{figure}
\begin{center}
 \includegraphics[width=0.9\columnwidth]{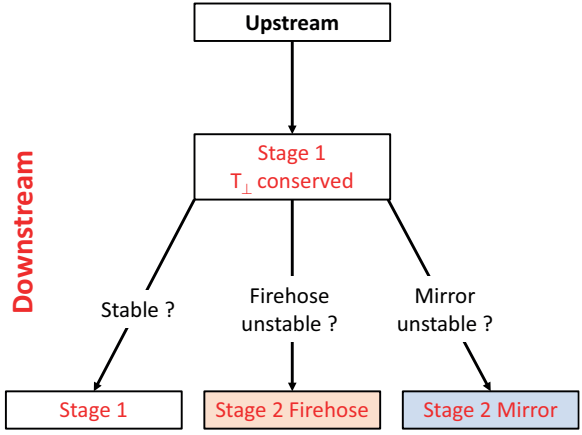}
\end{center}
\caption{Flowchart of the method explained here. As it crosses the chock front, the plasma conserves its temperature perpendicular to the motion. Here, this is $T_\parallel$. The resulting downstream is labelled ``Stage 1’’. If Stage 1 is stable, this is the end state of the downstream. If it is firehose unstable, the plasma migrates to the firehose marginal stability threshold. This is ``Stage 2-firehose’’. If it is mirror unstable, the plasma migrates to the mirror marginal stability threshold. This is ``Stage 2-mirror’’.  The color code is coherent with that of figure \ref{fig:bilan}.}\label{fig:flow}
\end{figure}

The method implemented can eventually be summarized by the flowchart pictured on figure \ref{fig:flow}.
\begin{itemize}
  \item As it crosses the front, the plasma has its $T_\parallel$ conserved. This results in a downstream labelled ``Stage 1''.
  \item If Stage 1 is stable, then this is the end state of the downstream.
  \item If Stage 1 is firehose unstable, the plasma migrates to the firehose marginal stability threshold. This is ``Stage 2-firehose'', end state of the downstream.
  \item If Stage 1 is mirror unstable, the plasma migrates to the mirror marginal stability threshold. This is ``Stage 2-mirror'', end state of the downstream.
\end{itemize}

It turns out that the stability of Stage 1 is a key ingredient of the model. We now elaborate on the instabilities involved, namely the firehose and the mirror instabilities.

\begin{figure}
\begin{center}
 \includegraphics[width=\columnwidth]{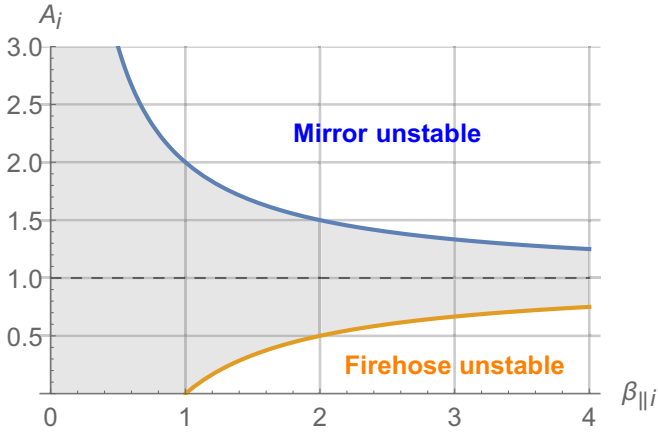}
\end{center}
\caption{Stability diagram. The plasma is stable within the shaded area.}\label{fig:stab}
\end{figure}

\subsection{Instabilities involved}\label{sec:insta}
For an homogenous plasma within a magnetic field $\bmath{B}_i$, stability relies on 2 parameters \citep{Hasegawa1975,Gary1993}. The anisotropy parameter,
\begin{equation}
A_i = \frac{T_{\perp i}}{T_{\parallel i}},
\end{equation}
where $i=1,2$, and the $\beta$ parameter,
\begin{equation}
\beta_i = \frac{  n_i k_B T_{\parallel i}}{B_i^2/8 \pi }.
\end{equation}
Figure \ref{fig:stab} pictures the stability diagram in the plane $(\beta_i,A_i)$. If
\begin{equation}\label{eq:mirror}
A_i > 1+\frac{1}{\beta_i},
\end{equation}
the plasma is mirror unstable. If
\begin{equation}\label{eq:firehose}
A_i < 1-\frac{1}{\beta_i},
\end{equation}
the plasma is firehose unstable. As evidenced, the topology is such that the plasma cannot be both mirror and firehose unstable at the same time.
Either it has $A_i<1$ and can be firehose unstable, or it has $A_i>1$ and can be mirror unstable.

Note that although the aforementioned criteria have been derived for electon/ion plasma, they are the same for pair plasmas \citep{Gary2009,Schlickeiser2010}.

\section{Results}\label{sec:results}
The model arising from implementing the steps described on figure \ref{fig:flow} can be solved exactly. For clarity, we report its main features here while the algebra is explained in Appendices.
\begin{itemize}
  \item To start with, there are limitations on $A_1$. The upstream cannot be so anisotropic that it is unstable. This sets limits on the minimum field parameter $\sigma_{m1}$ that can be considered for a given combination $(\chi_1,A_1)$, namely,
      \begin{equation}\label{eq:sigmaupstab}
  \sigma_{m 1}  \equiv \left\{
  \begin{array}{l}
   \frac{2}{\chi_1^2}(1-A_1), ~ \mathrm{when} ~ A_1 < 1 ~(\mathrm{firehose} ~ \mathrm{stability}), \\
  \frac{2}{\chi_1^2}(A_1-1), ~ \mathrm{when} ~ A_1 > 1 ~(\mathrm{mirror} ~ \mathrm{stability}).
  \end{array}
  \right.
   \end{equation}
      See Appendix \ref{sec:upstab} for the derivation.
  \item Generally speaking, the density jump at low $\sigma$ is given by Stage 2, because  Stage 1 is unstable at low $\sigma$. Indeed, the upstream resulting from the conservation of $T_\parallel$ is often\footnote{Not always. See below.} too anisotropic to be stabilized by a small field.
  \item At large $\sigma$, the density jump is given by Stage 1. The result, notably independent of the upstream anisotropy $A_1$ and derived in Appendix \ref{sec:rS1}, reads
  \begin{equation}\label{eq:rS1}
      r = \frac{3 \chi_1^2}{(2 \sigma +1) \chi_1^2+4}.
\end{equation}
  \item The transition from Stage 2 at low $\sigma$ to Stage 1 at high $\sigma$ occurs for a critical value of $\sigma$ below which Stage 1 is unstable.
  \item This critical value of $\sigma$ is a function of $\chi_1$ and $A_1$. The same parameters $\chi_1$ and $A_1$ also determine whether Stage 1 is firehose or mirror unstable.
  \begin{itemize}
      \item Stage 1 is firehose unstable for,
       \begin{equation}
          \sigma < \sigma_{fire} \equiv \frac{7}{4}-\frac{8 + 3 \chi_1 \sqrt{9 \chi_1^2-16/A_1}}{4 \chi_1^2}.
        \end{equation}
       See Appendix \ref{sec:S1firehose} for the derivation.
      \item Stage 1 is mirror unstable for  $\sigma < \sigma_{mirror}$. See Appendix \ref{sec:S1mirror} for, the expression of $\sigma_{mirror}$.
   \end{itemize}
   \item Then,
     \begin{itemize}
       \item If Stage 1 is firehose unstable, it switches to Stage 2-firehose, with a density jump derived in Appendix \ref{sec:rS2fire}.
       \item If Stage 1 is mirror unstable, it switches to Stage 2-mirror, with a density jump derived in Appendix \ref{sec:rS2mirror}.
     \end{itemize}
  \item Finally, it can happen that the minimum field required to stabilize the upstream is already large enough to stabilize Stage 1. In such a case, the jump is given by Stage 1 for any $\sigma > \sigma_{m1}$.
\end{itemize}

\begin{figure}
\begin{center}
 \includegraphics[width=\columnwidth]{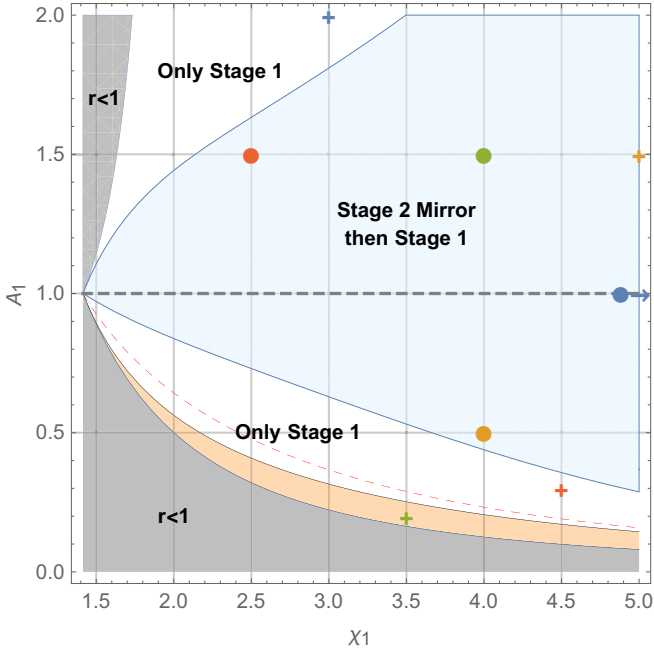}
\end{center}
\caption{Classification of the multiple kinds of solutions offered by the model in terms of $(\chi_1,A_1)$. In the orange region, the system goes from Stage 2-firehose at small $\sigma$, to Stage 1 at larger $\sigma$. The crosses show the cases for which the density jump is plotted as a function of $\sigma$ on figure \ref{fig:jumps}.  The circles show the cases for which the density jump is compared to the MHD result on figure \ref{fig:rS2mirrorstrong}. The color code of the regions is the same than that of figure \ref{fig:flow}.}\label{fig:bilan}
\end{figure}

The multiple kinds of solutions offered by the model can be classified on figure \ref{fig:bilan}, in terms of $(\chi_1,A_1)$ only. Some regions are simply discarded because they have $r<1$, which also entails a negative entropy jump, from the upstream to Stage 1 (see Appendix \ref{sec:rS1S}). In other regions, the minimum field needed to stabilize the upstream also stabilizes Stage 1. These are the regions labelled ``Only Stage 1''. The widest region, the blue one, is where the system goes from Stage 2-mirror to Stage 1 as $\sigma$ is increased. Finally, there is a tiny orange region, where it can go from Stage 2-firehose to Stage 1 with increasing $\sigma$. Its topology is better explained in Appendix \ref{sec:when}, figure \ref{fig:SigFireMirrorSigMin}.

\begin{figure}
\begin{center}
 \includegraphics[width=\columnwidth]{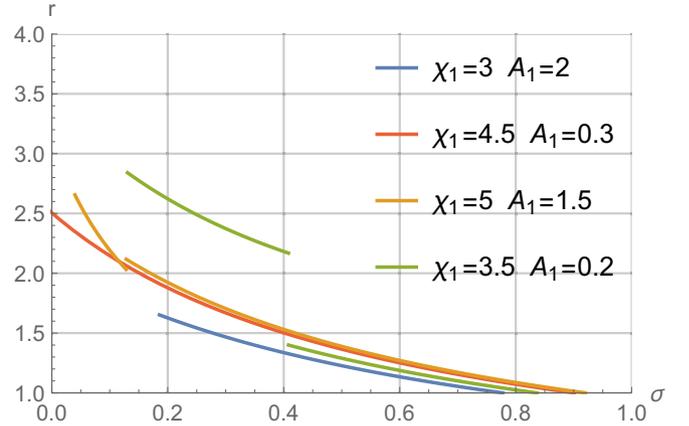}
\end{center}
\caption{Plots or $r(\sigma)$ for the parameters corresponding to the crosses on figure \ref{fig:bilan}.  The curves here and the corresponding crosses on figure \ref{fig:bilan} share the same color code.}\label{fig:jumps}
\end{figure}

Figure \ref{fig:jumps} shows the curves $r(\sigma)$ for the parameters corresponding to the crosses on figure \ref{fig:bilan}.

\begin{figure}
\begin{center}
 \includegraphics[width=\columnwidth]{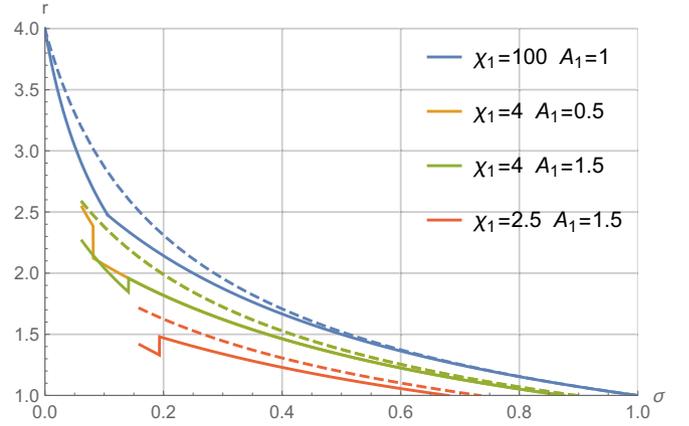}
\end{center}
\caption{Comparison between the MHD jump (dashed lines, Eq. \ref{eq:rMHD}) and the jump given by the present model (plain lines), for the parameters indicated by the circles on figure \ref{fig:bilan}. Same color code between the curves and the circles. In the limit $\chi_1 \rightarrow\infty$, the  result is independent of $A_1$. It is given by the solution of (\ref{eq:rS2mirrorstrong})  before the break (\ref{eq:breakstrong}), and by Eq. (\ref{eq:rS1Strong}) after the break.  Neither the MHD jump nor Stage 1 (our Eq. (\ref{eq:rS1}), relevant after the $\sigma$-break) depend on $A_1$.}\label{fig:rS2mirrorstrong}
\end{figure}

\subsection{Strong sonic shock limit  and comparison with MHD}
Simpler results are available in the strong sonic shock limit, namely $\chi_1 \rightarrow\infty$.
\begin{itemize}
  \item The minimum value (\ref{eq:sigmaupstab}) of $\sigma$ needed to stabilize the upstream goes to 0 for any $A_1$.
  \item The density jump (\ref{eq:rS1}) for Stage 1 goes to
  \begin{equation}\label{eq:rS1Strong}
  r(\sigma)=\frac{3}{2\sigma+1}.
  \end{equation}
  \item The critical field required to firehose stabilize Stage 1 goes to $-1/2$, which means Stage 1 is always firehose stable. Therefore, the density jump in the orange region in figure \ref{fig:bilan} is given by Stage 1, for any $\sigma\in [0,1]$. However, this region becomes infinitesimally small when $\chi_1 \rightarrow\infty$.
  \item The critical field required to mirror stabilize Stage 1 goes to
  \begin{equation}\label{eq:breakstrong}
  \sigma_{mirror} = 1 +  \frac{3}{2} \left[ (\sqrt{2}-1)^{1/3}- (\sqrt{2}+1)^{1/3}\right] \sim 0.1059.
  \end{equation}
The expression of the density jump in this case can be derived setting to 0 the coefficient of $\chi_1^2$ in the polynomial (\ref{eq:polyrS2mirror}). The result remains a 3rd degree equation in $r$,
      \begin{equation}\label{eq:rS2mirrorstrong}
      r \left[5 (\sigma +2)-2 r ((r+2) \sigma +1)\right]-8 = 0,
      \end{equation}
      but remarkably universal, that is, independent of $A_1$.

      Note that for any given $A_1$, figure \ref{fig:bilan} shows that the limit $\chi_1 \rightarrow\infty$ always ends up in the blue region. The relevant solution is plotted on figure \ref{fig:rS2mirrorstrong} (curve $\chi_1 = 100$). It is given by the solution of (\ref{eq:rS2mirrorstrong})  before the break (\ref{eq:breakstrong}), and by Eq. (\ref{eq:rS1Strong}) after the break. Obviously, we recover the result of \cite{BretPoP2019} for the strong shock case.
\end{itemize}

Finally, it is interesting to compare the result of the present model with the MHD one, even though MHD cannot deal with anisotropies. The ``isotropic MHD'' density jump is given by \citep{fitzpatrick,BretPoP2019}
\begin{eqnarray}\label{eq:rMHD}
r_{MHD}&=&\frac{\sqrt{\mathcal{C}} -(5 \sigma +2) \chi_1^2-10}{2 \sigma  \chi_1^2},  \\
\mathcal{C} &=& (\sigma +2) (25 \sigma +2) \chi_1^4+20 (5 \sigma +2) \chi_1^2+100,   \nonumber
\end{eqnarray}
which, in the strong sonic shock limit $\chi_1 \rightarrow\infty$, reduces to
\begin{equation}
r_{MHD_\infty}=\frac{\sqrt{(\sigma +2) (25 \sigma +2)}- (5 \sigma + 2)}{2 \sigma }.
\end{equation}
The dashed lines on figure \ref{fig:rS2mirrorstrong} features this MHD solutions. The largest departure from MHD is always found at the break between Stage 2 and Stage 1. It is reported on Table \ref{table} for the cases plotted on figure \ref{fig:rS2mirrorstrong}. The departure tends to be larger at small $\chi_1$, i.e. weak sonic shocks.

\begin{table}
\begin{tabular}{|c|c|c|c|}
  \hline
  $\chi_1=\infty$ & $\chi_1=4$, $A_1=0.5$  & $\chi_1=4$, $A_1=1.5$  & $\chi_1=2.5$, $A_1=1.5$  \\
  -12.5\%         & -14\%                  & -16\%       & -19\% \\
  \hline
\end{tabular}
\caption{ Maximum relative departure $(r_{MHD}-r)/r_{MHD}$, of the density jump $r$ given by the present model with respect to the MHD value $r_{MHD}$, for the cases displayed on figure \ref{fig:rS2mirrorstrong}. In terms of $\sigma$, the maximum departure always occurs at the break between Stage 2 and Stage 1.}\label{table}
\end{table}

\section{Conclusions}
In this article, we have applied a model previously developed, to compute the density jump of a collisionless perpendicular shock, when its upstream is anisotropic. The MHD formalism cannot achieve such a task for it lacks one conservation equation. Assuming the temperature perpendicular to the motion is conserved at the front crossing, these equations can be solved. This temperature conservation stems for the fact that the plasma is compressed in the direction parallel to the flow.

The state of the downstream resulting from such temperature conservation is called ``Stage 1''. Depending on the field strength, Stage 1 can be stable or not. At low field parameter $\sigma$, it is firehose or mirror unstable and migrates to the corresponding stability threshold. This is ``Stage2'' -mirror or -firehose, end state of the downstream. If the field is strong  enough, then Stage 1 can be stable, and be the end state of the downstream.

In \cite{BretMNRAS2023}, the case of a shock where the field is parallel to the direction of propagation was treated (``parallel shock''). Here, we treated the case of a field perpendicular to the direction of motion (``perpendicular shock''). It is interesting to compare both results.

For a parallel shock, our rule of temperature conservation at the front crossing results in a Stage 1 which tends to be firehose unstable. Yet, as evidenced on figure 5 of \cite{BretMNRAS2023}, for a high enough upstream anisotropy $A_1$, Stage 1 can also be mirror unstable.

For the present case of a perpendicular chock, our rule of temperature conservation at the front crossing results in a Stage 1 which tends to be mirror unstable. Yet, Stage 1 can also be firehose unstable, but now in a narrow region of the $(\chi_1,A_1)$ phase space, as evidenced by the orange region of figure \ref{fig:bilan} of the present article.

For both parallel and perpendicular shocks, the case $A_1=1$ is asymptotically recovered in the strong sonic shock limit, namely $\chi_1 \rightarrow \infty$. In this limit, it then becomes relevant to compare our model to the MHD result. For the present perpendicular case, this is done on figure \ref{fig:rS2mirrorstrong}. For the parallel case, one needs to compare the green curve of figure 6(c) of \cite{BretMNRAS2023}, with the MHD jump $r=4$. The departure from MHD is much more pronounced for the parallel case since there, within the MHD picture, the field does not have any influence on the density jump \citep{Lichnerowicz1976,Majorana1987}.

Note that the exploration of every possible field obliquities revealed that the largest departure from MHD is indeed found for parallel shocks \citep{BretJPP2022a}.

Direct comparisons with PIC simulations can be fruitful to further test the model presented, as was initiated in \cite{Haggerty2022}. Comparisons with MHD are only meaningful for $A_1=1$. For such a value of the $A_1$ parameter, PIC simulations may not be accurate enough to evidence differences between our model and the MHD behavior, as evidenced in figure \ref{fig:rS2mirrorstrong} for example. However, the present model makes predictions where MHD cannot, namely for $A_1 \neq 1$. There, PIC tests may prove enriching.

\section*{Acknowledgements}
A.B. acknowledges support by Grant  PID2021-125550OB-I00 from the Spanish Ministerio de Ciencia e Innovación.

%%%%%%%%%%%%%%%%%%%%%%%%%%%%%%%%%%%%%%%%%%%%%%%%%%
\section*{Data Availability}
The calculations presented in this manuscript were performed using \emph{Mathematica}. The Notebook files will be shared on
reasonable request to the corresponding author.



\bibliographystyle{mnras}
\bibliography{BibBret}

%\begin{thebibliography}{99}
%\bibitem[\protect\citeauthoryear{Author}{2013}]{author2013}
%\bibitem[\protect\citeauthoryear{Jones}{2015}]{jones2015}
%Jones C.~D., 2015, Journal of Interesting Stuff, 17, 198
%\bibitem[\protect\citeauthoryear{Smith}{2014}]{smith2014}
%Smith A.~B., 2014, The Example Journal, 12, 345 (Paper I)
%\end{thebibliography}

%%%%%%%%%%%%%%%%%%%%%%%%%%%%%%%%%%%%%%%%%%%%%%%%%%

%%%%%%%%%%%%%%%%% APPENDICES %%%%%%%%%%%%%%%%%%%%%

\appendix
\section{Restrictions on the upstream parameters}\label{sec:upstab}
Even though we consider the possibility of an anisotropic upstream, it still needs to be stable. This means its $A_1$ and $\beta_1$ parameters must fulfill both (\ref{eq:mirror}) and (\ref{eq:firehose}). It is convenient to recast these 2 conditions in terms of the field strength by reading the stability diagram on figure \ref{fig:stab} ``horizontally'', so to speak. With this in mind, upstream stability requires,
\begin{itemize}
  \item If $A_1 > 1$, then $\beta_1 < 1/(A_1-1)$ for firehose stability.
  \item If $A_1 < 1$, then $\beta_1 < 1/(1-A_1)$ for mirror stability.
\end{itemize}
In terms of the dimensionless variables (\ref{eqs:dimless}), these two equations allow to define a minimum value of the field parameter $\sigma_{m 1}$ necessary to have a stable upstream,
\begin{equation}\label{eq:sigmaupstabapp}
  \sigma_{m 1}  \equiv \left\{
  \begin{array}{l}
   \frac{2}{\chi_1^2}(1-A_1), ~ \mathrm{when} ~ A_1 < 1 ~(\mathrm{firehose} ~ \mathrm{stability}), \\
  \frac{2}{\chi_1^2}(A_1-1), ~ \mathrm{when} ~ A_1 > 1 ~(\mathrm{mirror} ~ \mathrm{stability}).
\end{array}
\right.
\end{equation}

\section{Study of Stage 1}\label{sec:rS1}
\subsection{Density and entropy jumps for Stage 1}\label{sec:rS1S}
We here study the properties of Stage 1, where the system (\ref{eq:conser1}-\ref{eq:conser4}) is closed with $T_{\parallel 2}=T_{\parallel 1}$. It is solved by using Eqs. (\ref{eq:conser1},\ref{eq:conser3}) to eliminate $v_2$ and $B_2$ everywhere. This then allows to eliminate $T_{\perp 2}$ between (\ref{eq:conser2}) and (\ref{eq:conser4}) and find an equation for the density jump $r$,
\begin{equation}
(r-1) \left[r \left((2 \sigma +1) \chi_1^2+4\right)-3 \chi_1^2\right] = 0,
\end{equation}
with solutions,
\begin{eqnarray}
r &=& 1, \nonumber\\
r &=& \frac{3 \chi_1^2}{(2 \sigma +1) \chi_1^2+4}, \label{eq:rS1nontriv}
\end{eqnarray}
which is the very same result found in \cite{BretPoP2019} for the case of a perpendicular chock with $A_1=1$. The jump of the non trivial solution (\ref{eq:rS1nontriv}) is larger than unity for,
\begin{equation}\label{eq:domain}
\sigma < \frac{\chi_1^2-2}{\chi_1^2}.
\end{equation}

It can be checked that this condition is equivalent to requesting a positive entropy jump for Stage 1. For a Maxwellian of the type
\begin{equation}
F= \frac{n}{\pi ^{3/2} \sqrt{a} b}  \exp \left(-\frac{v_y^2}{a}\right) \exp \left(-\frac{v_x^2+v_z^2}{b}\right),
\end{equation}
where $a = 2 k_B T_\parallel / m$ and $b = 2 k_B T_\perp / m$. The entropy density reads,
\begin{equation}
S = -k_B \int F \ln F d^3v = \frac{1}{2} k_B n \left[\ln \left(\pi ^3 a b^2\right)-2 \ln n+3\right].
\end{equation}
So that
\begin{eqnarray}
\Delta s &\equiv & \frac{S_2}{n_2}-\frac{S_1}{n_1}=\frac{k_B}{2}\left[ \ln \left(\frac{A_2^2 T_{\parallel 2}^3}{A_1^2 T_{\parallel 1}^3 } \right)-2 \ln r \right] \nonumber \\
&=& k_B \ln \left( \frac{1}{r} \frac{A_2}{A_1}  \right),
\end{eqnarray}
where conservation of $T_\parallel$ has been used. Replacing $r$ and $A_2$ by their expressions (\ref{eq:rS1nontriv},\ref{eq:A2S1})  and requesting $\Delta s>0$ yields exactly (\ref{eq:domain}).

\begin{figure}
\begin{center}
 \includegraphics[width=\columnwidth]{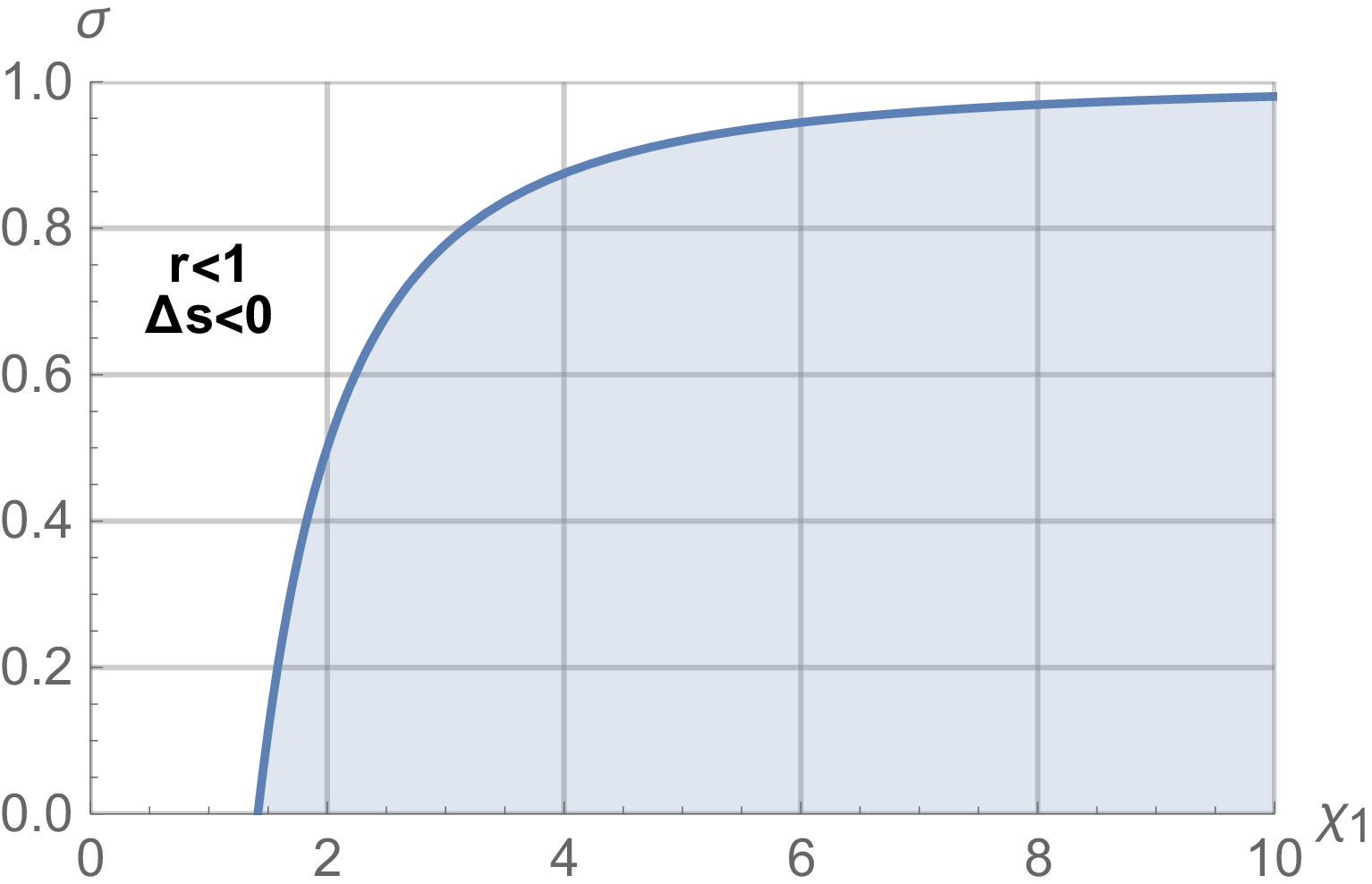}
\end{center}
\caption{The blue shaded area pictures the region of the $(\chi_1,\sigma)$ plane defined by  Eq. (\ref{eq:domain}), where the density jump for Stage 1 is larger than unity, and where the entropy jump is positive. The upstream anisotropy $A_1$ does not interfere.}\label{fig:rsup1}
\end{figure}

The domain of the $(\chi_1,\sigma)$ plane where condition (\ref{eq:domain}) is fulfilled is pictured on figure \ref{fig:rsup1} by the blue shaded area. Globally, $\chi_1$ is restricted to $\chi_1 > \sqrt{2}$, and $\sigma$ to $\sigma < 1$.

\subsection{Anisotropy and $\beta$ parameter Stage 1}\label{sec:A1beta1}
Since the switch to Stage 2 relies on the stability of Stage 1, which in turns relies on its anisotropy, we need now assess $A_2$ and $\beta_2$ in Stage 1. For $A_2$ we find

\begin{equation}\label{eq:A2S1}
A_2= A_1 \frac{\chi_1^2 \left[r^3 (-\sigma )+r (\sigma +2)-2\right]+2 r}{2 r^2},
\end{equation}
which is the result of \cite{BretPoP2019}, times $A_1$.

\begin{figure}
\begin{center}
 \includegraphics[width=\columnwidth]{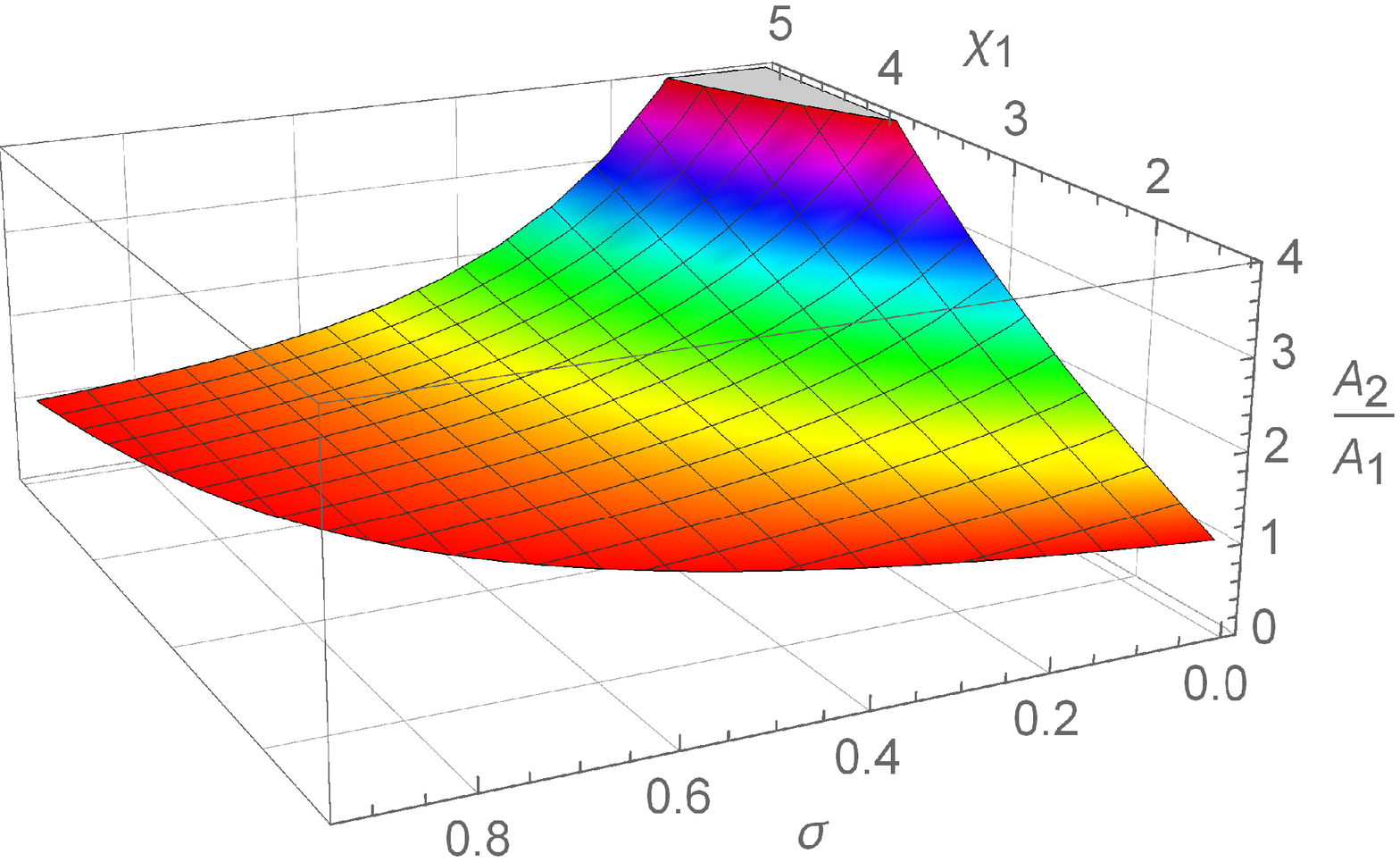}
\end{center}
\caption{Plot of Eq. (\ref{eq:A2S1}) over the domain defined by (\ref{eq:domain}).}\label{fig:S1A2}
\end{figure}

The quantity $A_2/A_1$ is plotted on figure \ref{fig:S1A2}.
\begin{itemize}
  \item For $A_1 = 1$, we find $A_2 > 1$ always, so that Stage 1 can only be mirror unstable.
  \item For $A_1 > 1$, $A_2 > 1$ so that Stage 1, if unstable, can only be mirror unstable.
  \item But for $A_1<1$, Stage 1 can be both mirror \emph{and} firehose unstable.
\end{itemize}

Then we explain the $\beta_2$ parameter,
\begin{equation}
\beta_2 = \frac{  n_2k_B T_{\parallel 2}}{B_2^2/8 \pi }.
\end{equation}
Since in Stage 1, $T_{\parallel}$ is conserved, we set $T_{\parallel 2}=T_{\parallel 1}$ and $\beta_2$ reads,
\begin{equation}
\beta_2 = \frac{ n_2k_B T_{\parallel 1}}{B_2^2/8 \pi} = \frac{2}{A_1 r \sigma  \chi_1^2},
\end{equation}
with $r$ for Stage 1 given by Eq. (\ref{eq:rS1nontriv}).

\begin{figure}
\begin{center}
 \includegraphics[width=0.49\columnwidth]{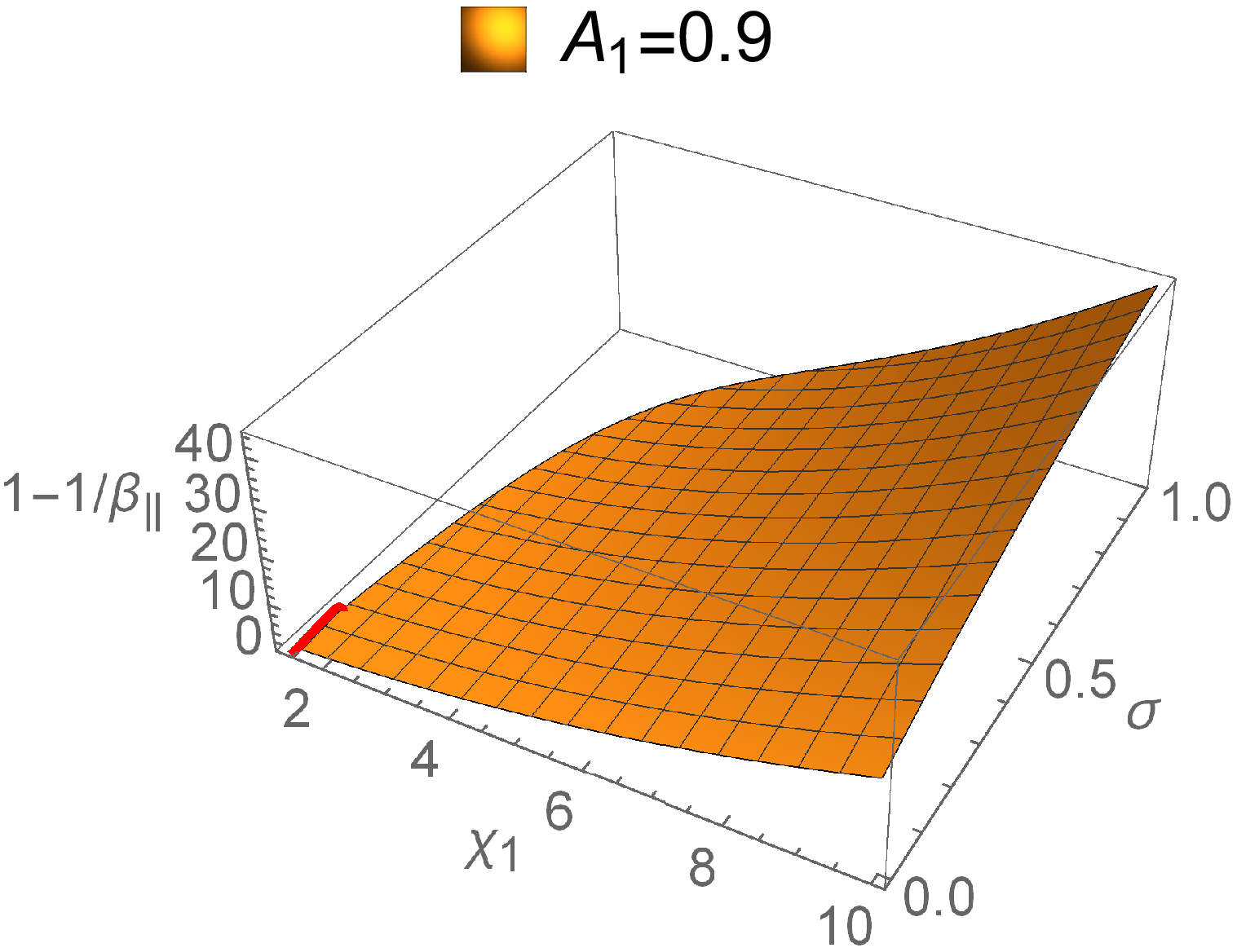} \includegraphics[width=0.49\columnwidth]{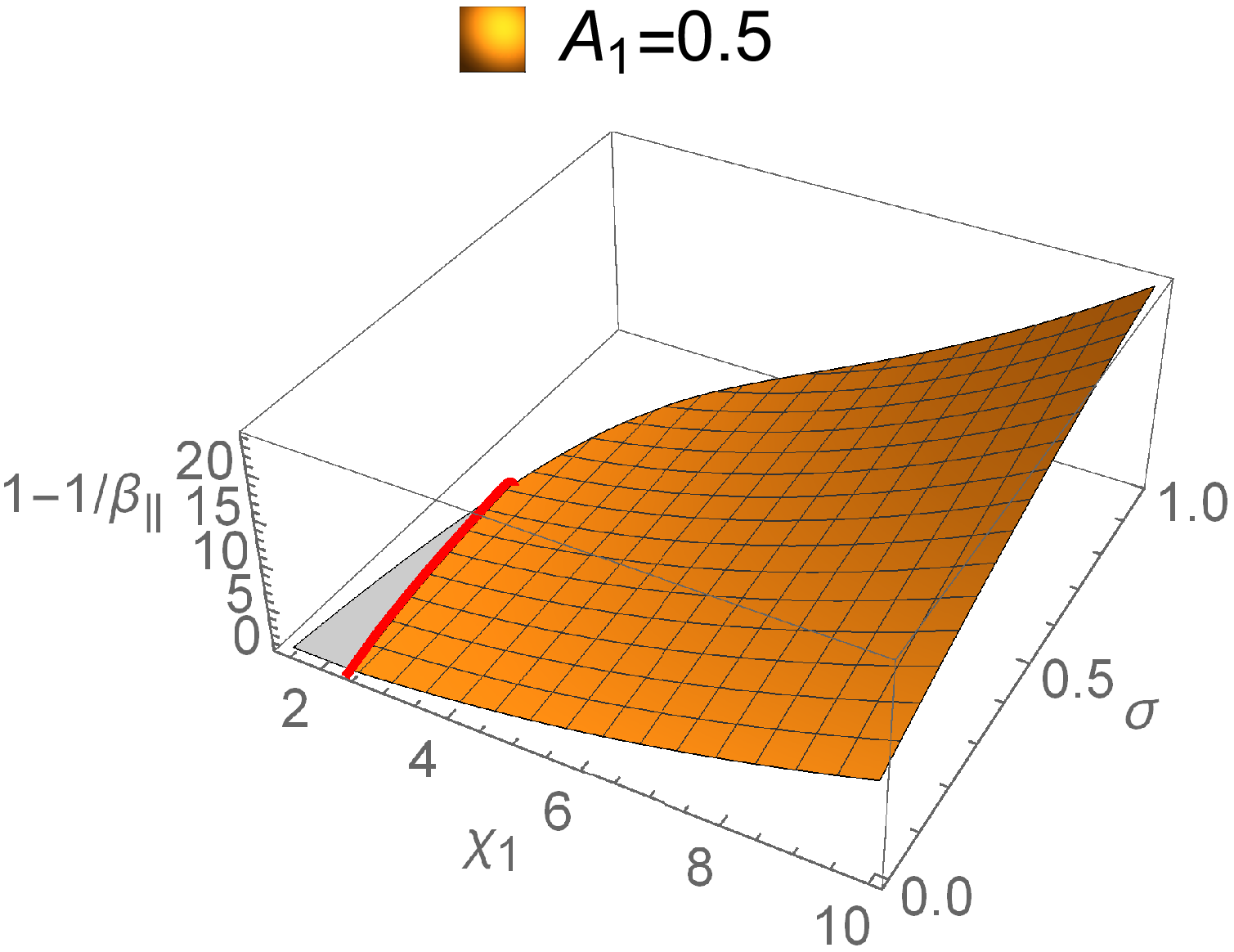}\\
  \includegraphics[width=0.49\columnwidth]{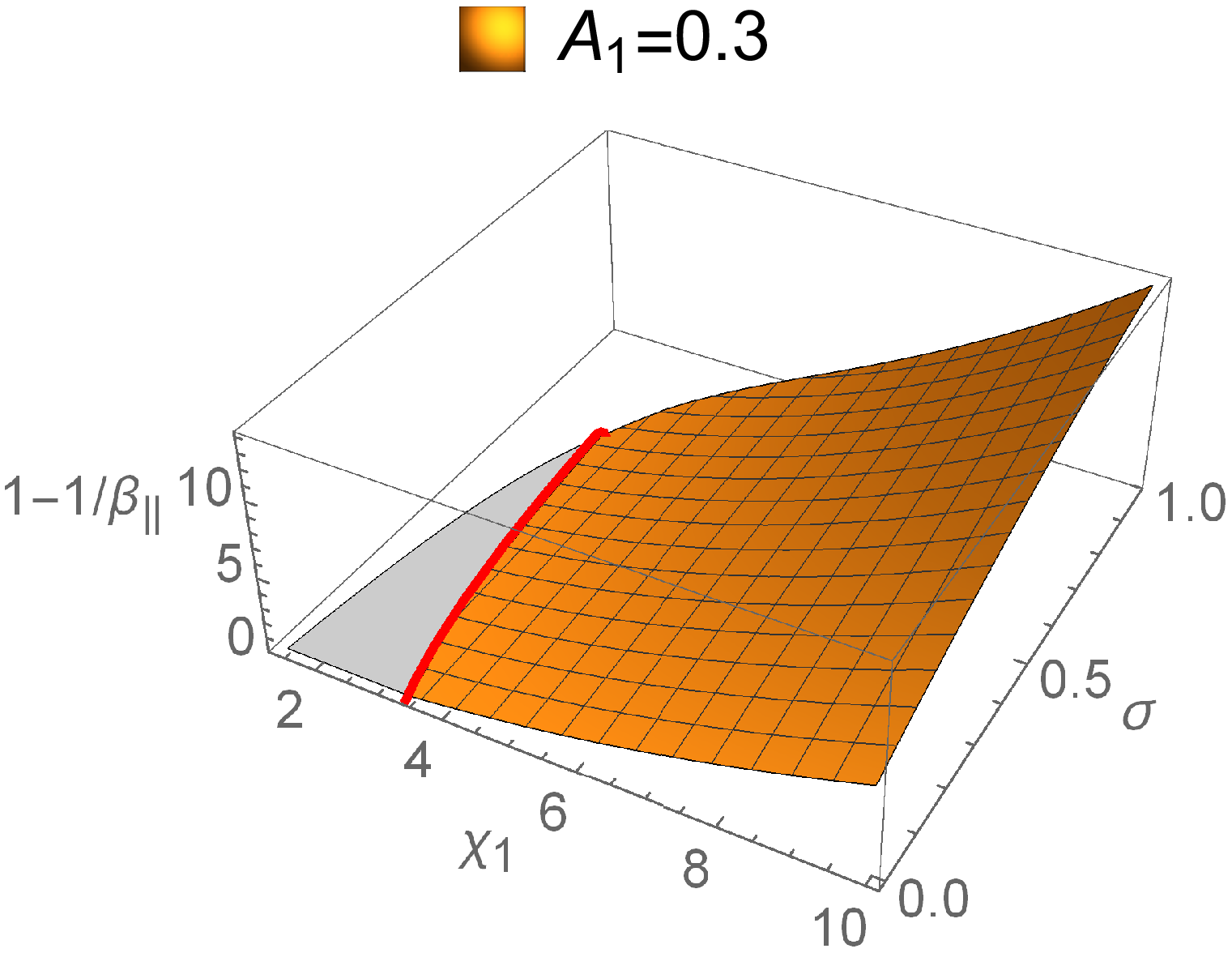} \includegraphics[width=0.49\columnwidth]{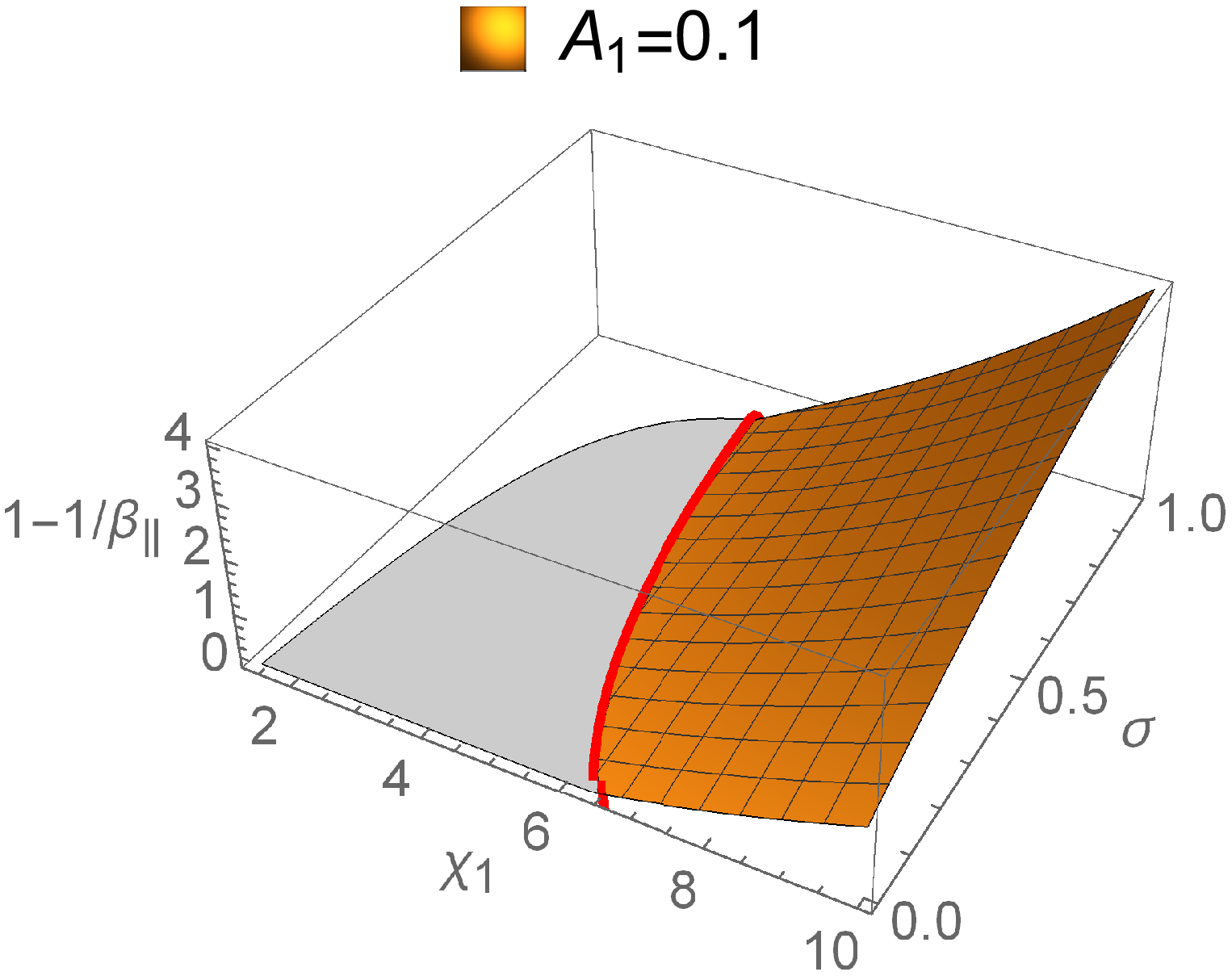}
\end{center}
\caption{Quantity $A_2-(1 - 1/\beta_2)$ for Stage 1. Stage 1 is firehose stable when this quantity is positive. The red line pictures the threshold given by Eq. (\ref{eq:sigcritfire}).}\label{fig:fireS1}
\end{figure}

\subsection{Firehose stability of Stage 1}\label{sec:S1firehose}
Stage 1 is marginally firehose stable if,
\begin{equation}
A_1 = 1 - \frac{1}{\beta_2}.
\end{equation}
It is reached for,
\begin{equation}\label{eq:sigcritfire}
\sigma \equiv \sigma_{fire} = \frac{7}{4}-\frac{8 + 3 \chi_1 \sqrt{9 \chi_1^2-16/A_1}}{4 \chi_1^2},
\end{equation}
so that Stage 1 is firehose unstable if $\sigma < \sigma_{fire}$. The quantity $A_2-(1 - 1/\beta_2)$ for Stage 1 is plotted on figure \ref{fig:fireS1} for 4 values of $A_1$.  Stage 1 is firehose stable when this quantity is positive, that is, for $\sigma$ large enough.  The red line pictures the threshold given by Eq. (\ref{eq:sigcritfire}). For $A_1=1$, Stage 1 is never firehose unstable, as found in \cite{BretPoP2019}.

\begin{figure}
\begin{center}
 \includegraphics[width=0.49\columnwidth]{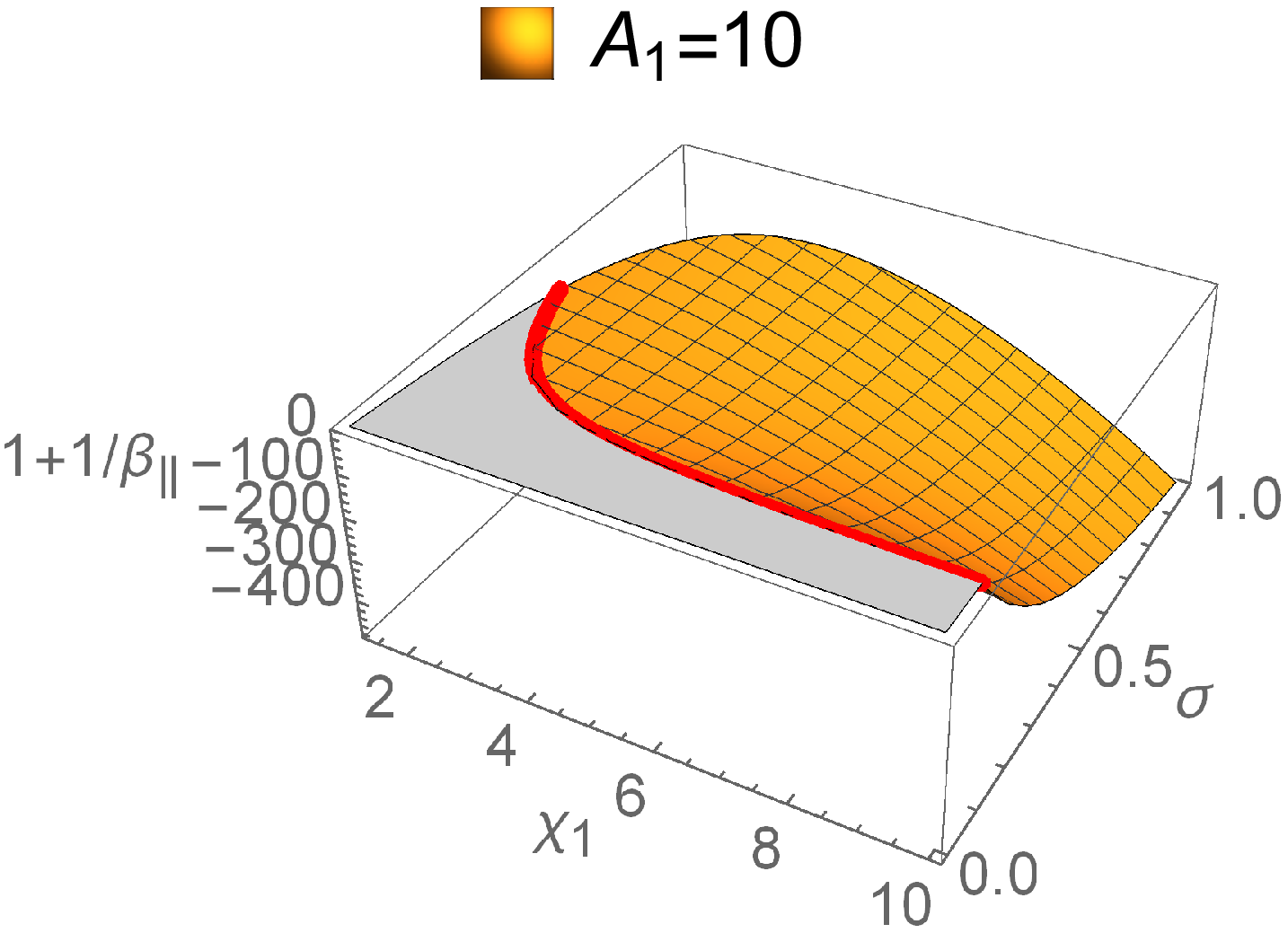} \includegraphics[width=0.49\columnwidth]{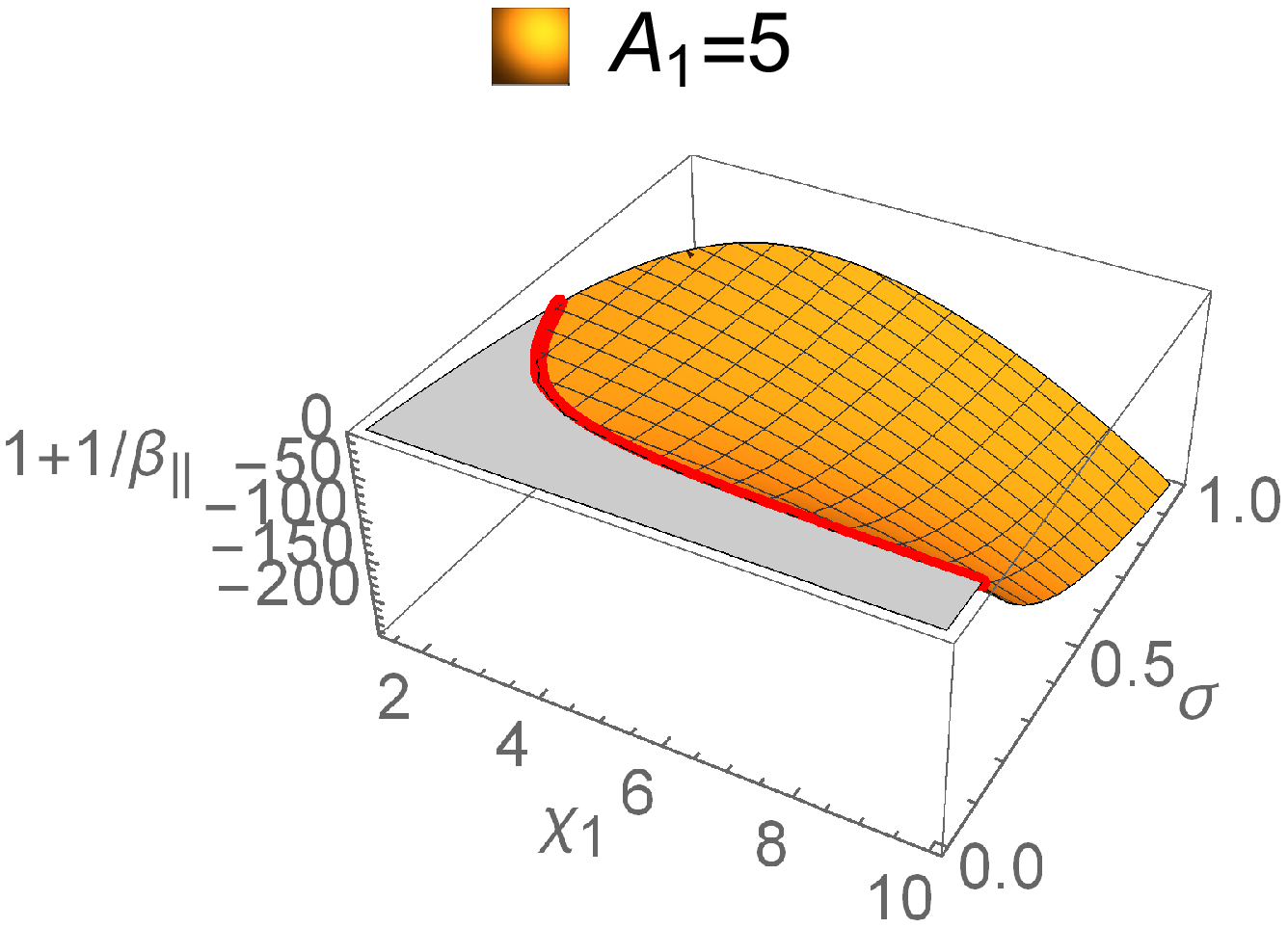}\\
  \includegraphics[width=0.49\columnwidth]{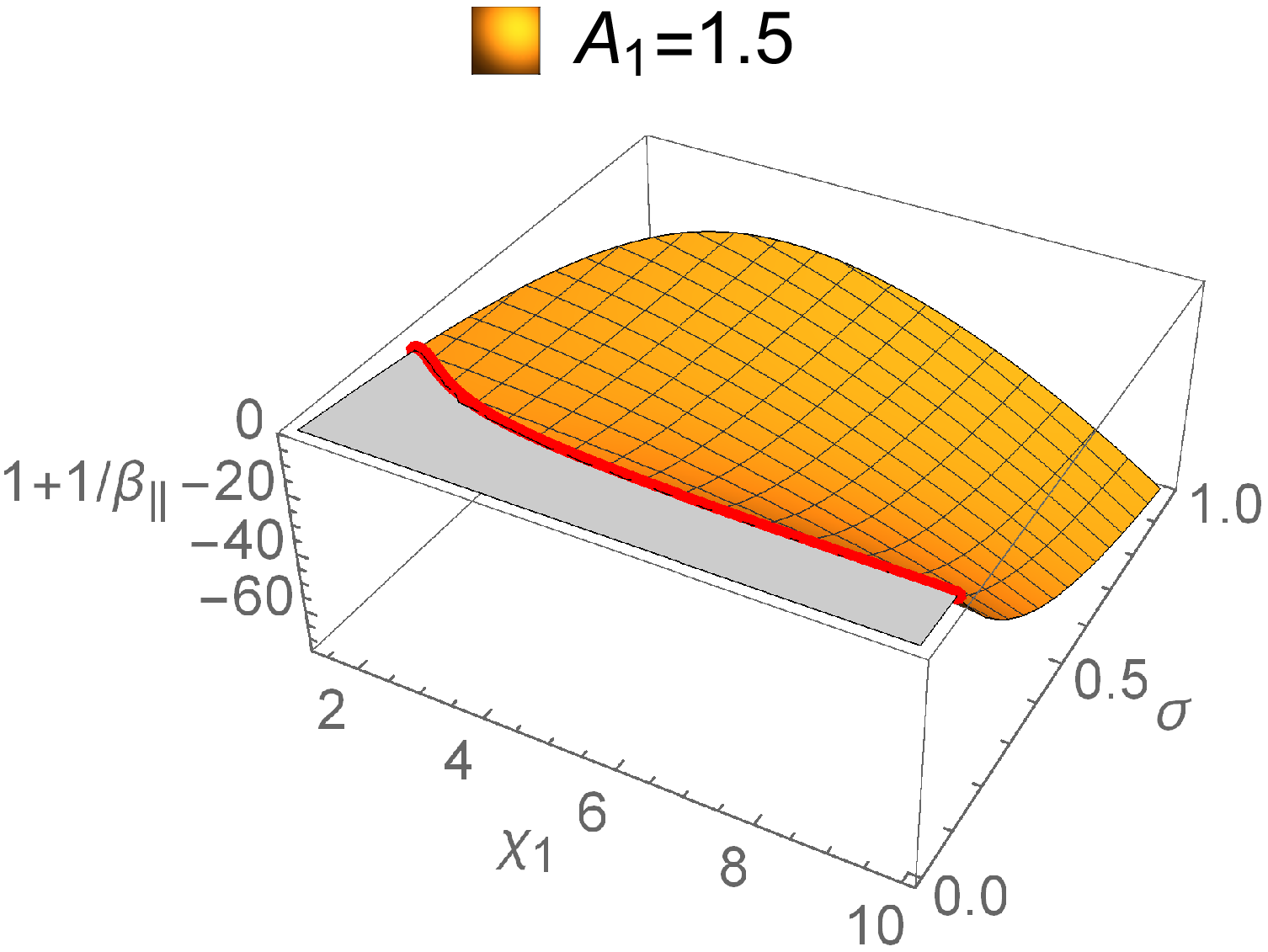} \includegraphics[width=0.49\columnwidth]{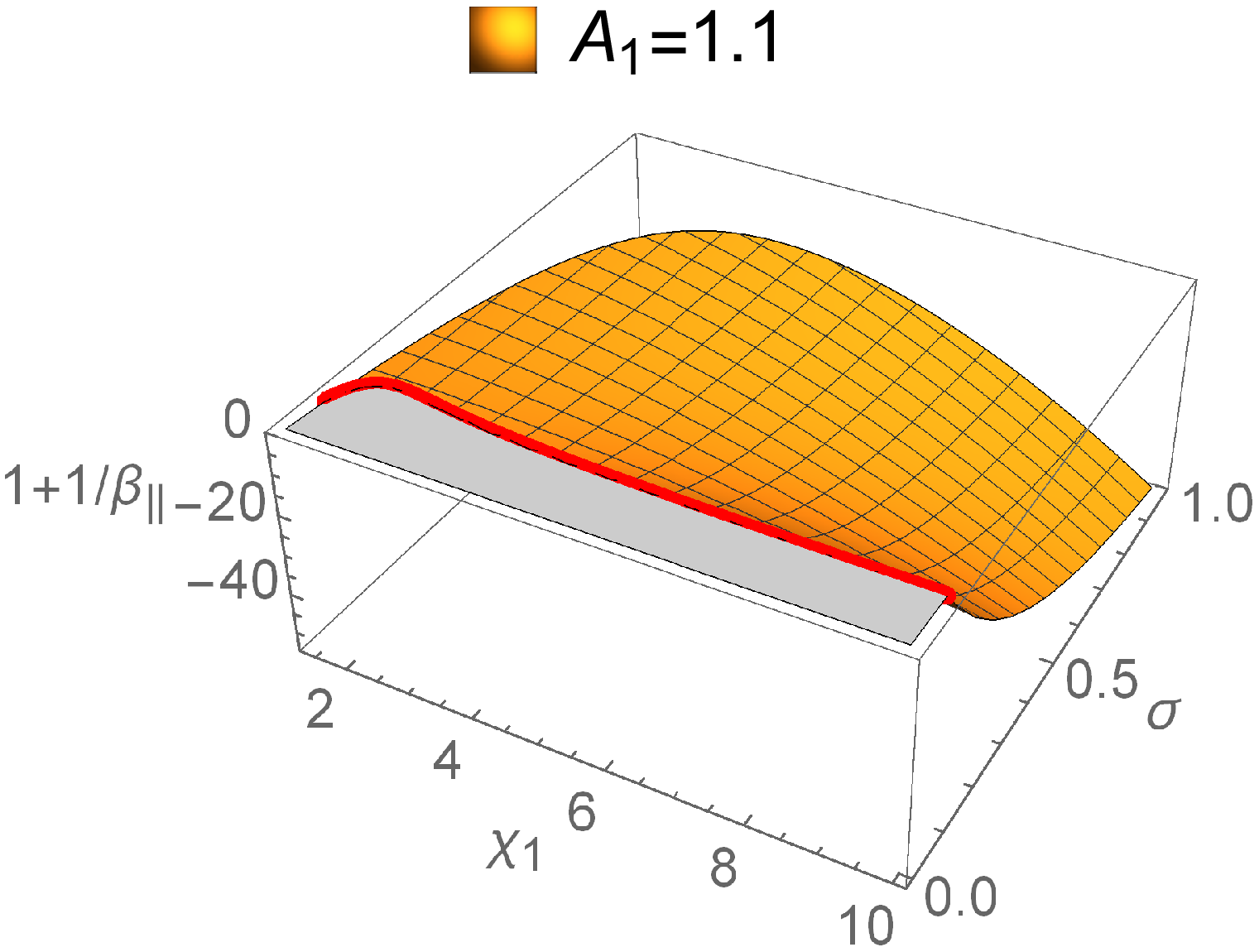}
\end{center}
\caption{Quantity $A_2-(1 + 1/\beta_2)$ for Stage 1. Stage 1 is mirror stable when this quantity  is negative. The red line pictures the threshold given by Eq. (\ref{eq:sigcritmirror}).}\label{fig:mirrorS1}
\end{figure}

\subsection{Mirror stability of Stage 1}\label{sec:S1mirror}

Stage 1 is marginally mirror stable if,
\begin{equation}\label{eq:S1margmirror}
A_1 = 1 + \frac{1}{\beta_2}.
\end{equation}
The quantity $A_2-(1 + 1/\beta_2)$ for Stage 1 is plotted on figure \ref{fig:mirrorS1} for 4 values of $A_1$.  Stage 1 is mirror stable when this quantity is negative, that is, for $\sigma$ large enough.

Solving Eq. (\ref{eq:S1margmirror})  gives a 3rd order polynomial for $\sigma_{mirror}$ under which Stage 1 is mirror unstable. The solution corresponding to the threshold we are interested in reads,
\begin{equation}\label{eq:sigcritmirror}
\sigma \equiv \sigma_{mirror} = \frac{3^{2/3} \left(3 A_1 \chi_1^2+4\right)}{\mathcal{A}}-\frac{\mathcal{B}}{2 A_1 \chi_1}-\frac{2}{\chi_1^2}+1,
\end{equation}
where,
\begin{eqnarray}
   \mathcal{A} &=& 2 \chi_1 \left[9 A_1^2 \chi_1 \left(A_1 (\chi_1^2-4)+2\right)+ \sqrt{6\Delta}\right]^{1/3}, \nonumber\\
   \mathcal{B} &=&  \left[27 A_1^2 \chi_1 \left(A_1 (\chi_1^2-4)+2\right)+3 \sqrt{6\Delta}\right]^{1/3}  ,\\
   \Delta &=& A_1^3 \left[9 A_1 \chi_1^2 \left(3 A_1 \left(A_1 \left(\chi_1^4-4 \chi_1^2+8\right)+4 \left(\chi_1^2-2\right)\right)+14\right)+32\right].   \nonumber
\end{eqnarray}
Figure \ref{fig:mirrorS1} features this threshold as the bold red line.

\begin{figure}
\begin{center}
 \includegraphics[width=\columnwidth]{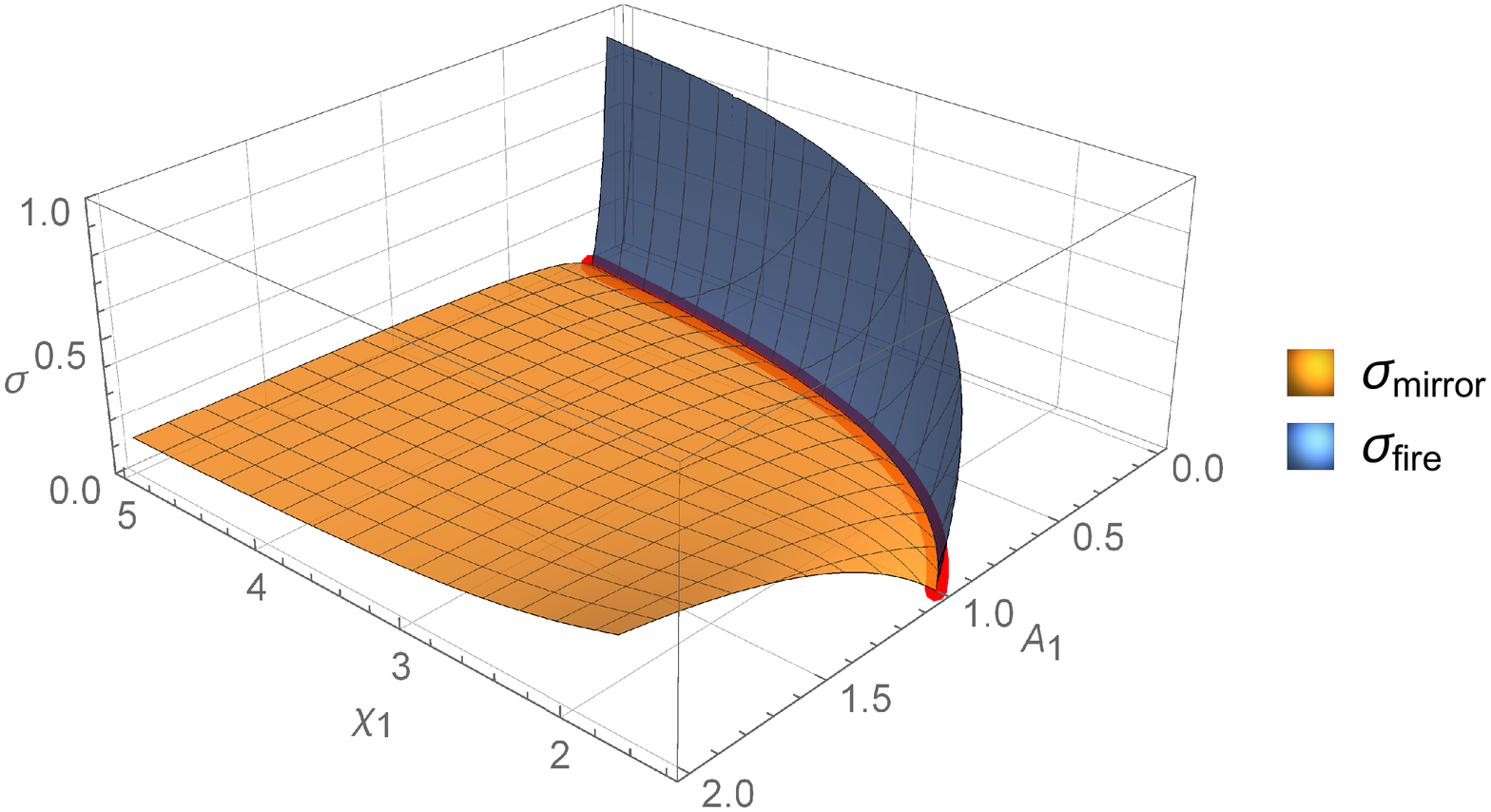}
\end{center}
\caption{Under the orange surface, Stage 1 is mirror unstable. Under the blue surface, it is firehose unstable. Above the 2 surfaces up to $\sigma=1$, the field is strong enough to stabilize Stage 1. The red line pertains to Eq. (\ref{fig:frontier}).}\label{fig:SigFireMirror}
\end{figure}

\begin{figure}
\begin{center}
 \includegraphics[width=\columnwidth]{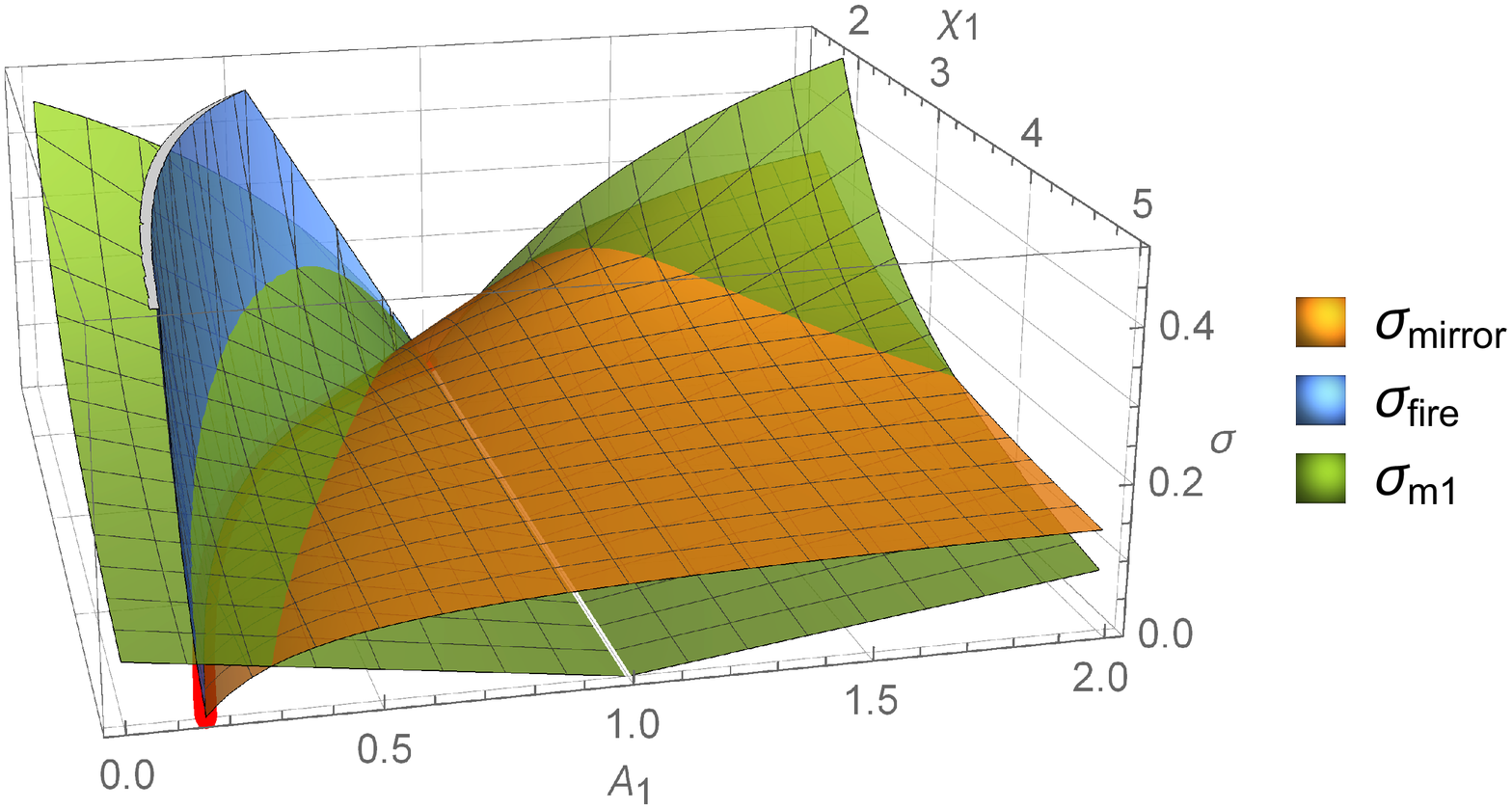}
\end{center}
\caption{Same as figure \ref{fig:SigFireMirror}, but viewed from the high $\chi_1$'s perspective, plus featuring $\sigma_{m1}$ from Eq. (\ref{eq:sigmaupstabapp}), the minimum $\sigma$ required to stabilize the upstream.}\label{fig:SigFireMirrorSigMin}
\end{figure}

\subsection{When can Stage 1 be firehose or mirror unstable?}\label{sec:when}
At this junction, it is interesting to determine when Stage 1 can be firehose or mirror unstable. For this purpose we now compare the threshold $\sigma_{fire}$ for firehose instability, with the threshold $\sigma_{mirror}$ for mirror instability. Plotting both quantities on the same plot yields figure \ref{fig:SigFireMirror}. As it appears, the $(\chi_1,A_1)$ plane is divided into 2 main regions,
\begin{itemize}
  \item One region where Stage 1 can be firehose unstable. This region has always $A_1 < 1$.
  \item One region where Stage 1 can be mirror unstable. This region has $A_1 < 1$ or $A_1>1$.
\end{itemize}
These 2 regions are separated by a frontier where $\sigma_{fire}=\sigma_{mirror}=0$. Solving from $\sigma_{fire}=0$ gives,
\begin{equation}\label{fig:frontier}
A_1=\frac{9 \chi_1^2}{2 \chi_1^4+7 \chi_1^2-4},
\end{equation}
which appears as the red line on figure \ref{fig:SigFireMirror}. Therefore, shocks located to the left of the red line can only have Stage 1 mirror unstable. Shocks located to the right of the red line can only have Stage 1 firehose unstable.

Figure \ref{fig:SigFireMirrorSigMin} is similar to figure \ref{fig:SigFireMirror}, but viewed from the high $\chi_1$'s perspective, and featuring $\sigma_{m1}$, the minimum $\sigma$ required to stabilize the upstream. It explains the intricate topology around the orange region in figure \ref{fig:bilan}.

\section{Density jump in Stage 2}\label{sec:rS2}
Having fully characterized Stage 1 and its stability, we may now turn to Stage 2. We only need to compute its density jump since by definition Stage 2 is (marginally) stable. The only distinction at this junction is between Stage 2-firehose, and Stage 2-mirror.

\subsection{Density ratio $r$ in Stage 2-firehose}\label{sec:rS2fire}
For Stage 1, we closed the system (\ref{eq:conser1}-\ref{eq:conser4}) setting $T_{\parallel 2}=T_{\parallel 1}$. We now close it setting $A_2=1-1/\beta_2$ instead. When solving the system (\ref{eq:conser1}-\ref{eq:conser4}) with this new prescription, the density jump $r$ is found to be solution of,
%
%\begin{equation}
%-2 r \left[A_1 (4 r-5)+(r-5) \chi_1^2+r\right]+r (5-4 r) \sigma  \chi_1^2-8 \chi_1^2 = 0
%\end{equation}
\begin{equation}
  \sum_{k=0}^2 rS2f_k r^k = 0,
\end{equation}
with,
\begin{eqnarray}
  rS2f_0 &=& -8 \chi_1^2, \\
  rS2f_1 &=& 5 \left(2 A_1+(\sigma +2) \chi_1^2\right), \nonumber\\
  rS2f_2 &=&  -2 \left(4 A_1+(2 \sigma +1) \chi_1^2+1\right). \nonumber
\end{eqnarray}
Here, one root has $r<1$ so that the other one is the physical one. It reads,
\begin{equation}\
r=\frac{1}{4}\frac{10 A_1+5 (\sigma +2) \chi_1^2 +\sqrt{\Delta }}{4 A_1+(2 \sigma +1) \chi_1^2+1},
\end{equation}
with
\begin{equation}
\Delta=100 A_1^2+4 A_1 (25 \sigma -14) \chi_1^2+\left[\sigma  (25 \sigma -28)+36\right] \chi_1^4-64 \chi_1^2.
\end{equation}

\subsection{Density ratio $r$ in Stage 2-mirror}\label{sec:rS2mirror}
We now close the system (\ref{eq:conser1}-\ref{eq:conser4}) setting $A_2=1+1/\beta_2$ instead. Here $r$ is found to be the solution of,
\begin{equation}\label{eq:polyrS2mirror}
  \sum_{k=0}^3 rS2m_k r^k =0,
\end{equation}
with,
\begin{eqnarray}
  rS2m_0 &=&   -8 \chi_1^2 , \\
  rS2m_1 &=&   5 \left(2 A_1+(\sigma +2) \chi_1^2\right) , \nonumber\\
  rS2m_2 &=&   -2 \left(4 A_1+(2 \sigma +1) \chi_1^2+1\right)  , \nonumber\\
  rS2m_3 &=&  -2 \sigma  \chi_1^2 . \nonumber
\end{eqnarray}
Out of the 3 roots of this equation, one has $r<0$ and another $A_2<0$. The remaining one is the physical one.

%%%%%%%%%%%%%%%%%%%%%%%%%%%%%%%%%%%%%%%%%%%%%%%%%%

% Don't change these lines
\bsp	% typesetting comment
\label{lastpage}
\end{document}